# New Alkali-Metal- and 2-Phenethylamine-Intercalated Superconductors $A_x(C_8H_{11}N)_yFe_{1-z}Se$ (A = Li, Na) with the Largest Interlayer Spacings and $T_c \sim 40$ K

Takehiro Hatakeda, Takashi Noji, Kazuki Sato, Takayuki Kawamata, Masatsune Kato, and Yoji Koike

*Department of Applied Physics, Tohoku University, 6-6-05 Aoba, Aramaki, Aoba-ku, Sendai 980-8579, Japan*

New FeSe-based intercalation superconductors, $A_x(C_8H_{11}N)_yFe_{1-z}Se$ (A = Li, Na), with $T_c = 39 - 44$ K have been successfully synthesized via the intercalation of alkali metals and 2-phenethylamine into FeSe. The interlayer spacings, namely, the distances between neighboring Fe layers, $d$, of $A_x(C_8H_{11}N)_yFe_{1-z}Se$ (A = Li, Na) are 19.04(6) and 18.0(1) Å, respectively. These $d$ values are the largest among those of the FeSe-based intercalation compounds and are understood to be due to the intercalation of two molecules of 2-phenethylamine in series perpendicular to the FeSe layers. It appears that the relationship between $T_c$ and $d$ in the FeSe-based intercalation superconductors is not domic but $T_c$ is saturated at ~ 45 K, which is comparable to the $T_c$ values of single-layer FeSe films, for $d \geq 9$ Å.

The compound FeSe has the simplest crystal structure among the iron-based superconductors and is composed of a stack of edge-sharing FeSe$_4$-tetrahedra layers (simply called FeSe layers). Although the superconducting transition temperature $T_c$ is only 8 K,[1] it markedly increases to ~ 45 K via the co-intercalation of alkali or alkaline-earth metals and ammonia or organic molecules between the FeSe layers. By collecting the data of $T_c$ and the interlayer spacing between neighboring Fe layers, $d$, of a variety of FeSe-based intercalation superconductors,[2-30] it has been found that the relationship between $T_c$ and $d$ appears domic.[16] That is, $T_c$ tends to increase with increasing $d$, have a maximum value of ~ 45 K at $d = 9 - 12$ Å, and then decrease. Recently, single-layer FeSe films on SrTiO$_3$ substrates have attracted considerable interest, because the opening of a possible superconducting gap has been observed at a temperature as high as 42 – 65 K in scanning tunneling microscopy/spectroscopy (STM/STS) [31] and angle-resolved photoemission spectroscopy (ARPES) [32,33] measurements. Moreover, electrical resistivity measurements have revealed a superconducting transition at ~ 40 K, although the transition is rather broad.[31,34,35] Since the single-layer FeSe films may be regarded as a kind of FeSe-based intercalation superconductor with infinite $d$ values, the electronic state and superconductivity of the single-layer FeSe films may be comparable to those of FeSe-based intercalation superconductors with large $d$ values.

Here, we report on the successful synthesis of new FeSe-based intercalation superconductors, $A_x(C_8H_{11}N)_y Fe_{1-z}Se$ ($A$ = Li, Na), with the largest $d$ values of ~ 19 Å and $T_c$ = 39 – 44 K via the co-intercalation of alkali metals and 2-phenethylamine (2-PEA), $C_8H_{11}N$, into FeSe. The relationship between $T_c$ and $d$ is discussed and compared with the $T_c$ values of the single-layer FeSe films.

Polycrystalline host samples of FeSe were prepared by the high-temperature solid-state reaction method as described in Ref. 15. Iron powder and selenium grains, which were weighed in a molar ratio of Fe:Se = 1.02:1, were mixed, put into an alumina crucible, and sealed in an evacuated quartz tube. This was heated at 1027 ℃ for 30 h and then annealed at 400 ℃ for 50 h, followed by furnace-cooling. The obtained ingot of FeSe was pulverized into powder to be used for the intercalation. Both the alkali metal and 2-PEA were intercalated into the powdery FeSe as follows. An appropriate amount of the powdery FeSe was placed in a beaker filled with a 0.2 M solution of pure lithium or sodium metal in 2-PEA. The amount of FeSe was calculated in the molar ratio of $A$:FeSe = 1:1. The reactions using lithium and sodium were carried out at 45 ℃ for 7 and 28 days, respectively. The product was washed with hexane. All the processes were performed in an argon-filled glove box.

Both the host sample of FeSe and the intercalated samples were characterized by powder x-ray diffraction using Cu K$\alpha$ radiation. For the intercalated samples, an airtight sample holder was used. The diffraction patterns were analyzed using RIETAN-FP.[36] Thermogravimetric (TG) measurements were performed in flowing gas of argon using a commercial analyzer (SII NanoTechnology Inc., TG/DTA7300). To detect the superconducting transition, the magnetic susceptibility $\chi$ was measured using a superconducting quantum interference device (SQUID) magnetometer (Quantum Design, MPMS).

Figure 1 shows the powder x-ray diffraction patterns of the host sample of FeSe and intercalated samples of $A_x(C_8H_{11}N)_yFe_{1-z}Se$ ($A$ = Li, Na). The broad peak around $2\theta = 20°$ is due to the airtight sample holder. Most of the sharp Bragg peaks are due to the intercalation compound of $A_x(C_8H_{11}N)_yFe_{1-z}Se$ ($A$ = Li, Na) and the host compound of FeSe, so that they can be indexed on the basis of the PbO-type ($P4/nmm$) structure.[37] It is found that a non-intercalated region of FeSe still remains in the intercalated samples. The $c$-axis lengths of $A_x(C_8H_{11}N)_yFe_{1-z}Se$ ($A$ = Li, Na) are calculated to be 19.04(6) and 18.0(1) Å, respectively, as listed in Table I. The $d$ value is the same as the $c$-axis length in the PbO-type structure.

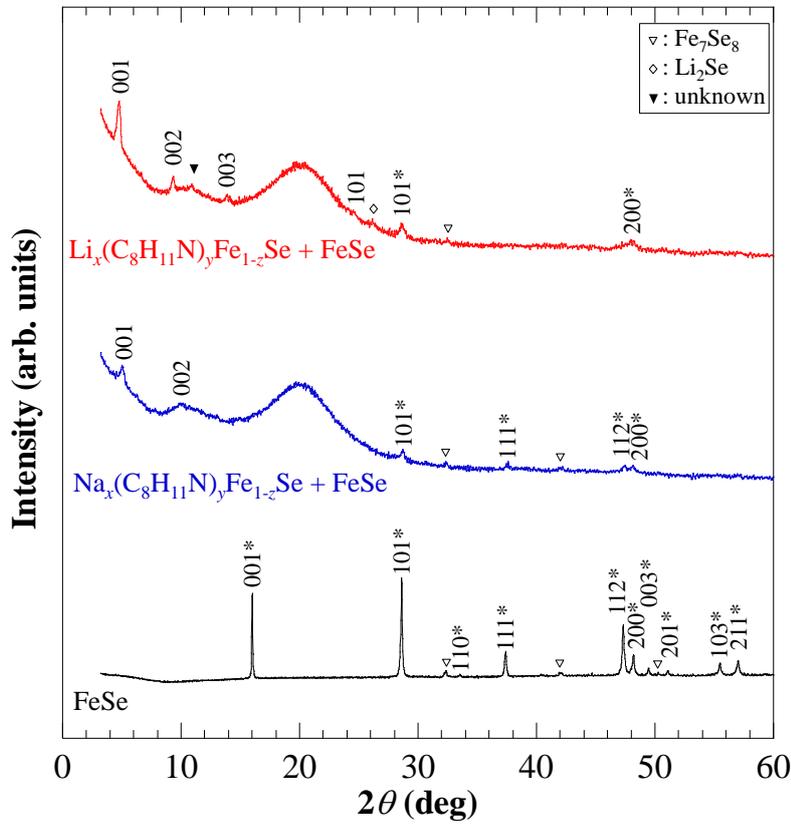

Fig. 1. Powder x-ray diffraction patterns of the host compound FeSe and the intercalated samples consisting of $A_x(C_8H_{11}N)_yFe_{1-z}Se$ ($A$ = Li, Na) and FeSe, obtained using Cu K$\alpha$ radiation. Indices without and with an asterisk are due to $A_x(C_8H_{11}N)_yFe_{1-z}Se$ ($A$ = Li, Na) and FeSe, respectively. All indices are based on the PbO-type ($P4/nmm$) structure. Peaks marked by ▽, ◇, and ▼ are due to $Fe_7Se_8$, $Li_2Se$, and an unknown compound, respectively. The broad peak around $2\theta = 20°$ is due to the airtight sample holder.

Table I. $c$-axis lengths of the host compound FeSe and $A_x(C_8H_{11}N)_yFe_{1-z}Se$ ($A$ = Li, Na) in the intercalated samples (in Å). It is noted that the distance between neighboring Fe layers, $d$, is the same as the $c$-axis length.

|  | $c = d$ |
|---|---|
| FeSe (host) | 5.504(3) |
| $Li_x(C_8H_{11}N)_yFe_{1-z}Se$ | 19.04(6) |
| $Na_x(C_8H_{11}N)_yFe_{1-z}Se$ | 18.0(1) |

Since the intercalation of only lithium into Fe(Se,Te) has no effect on either the superconductivity or the crystal structure,[4] it is concluded that not only lithium or sodium but also 2-PEA has been intercalated between the FeSe layers. The present $d$ values are the largest among those of the FeSe-based intercalation compounds. The enhancement of the $d$ value from that of FeSe to that of $A_x(C_8H_{11}N)_yFe_{1-z}Se$ ($A$ = Li, Na) is ~ 13 Å and about twice larger than the length of 2-PEA shown in Fig. 2(a). Therefore, it is inferred that two molecules of 2-PEA are intercalated in series perpendicular to the FeSe layers, as shown in Fig. 2(b).

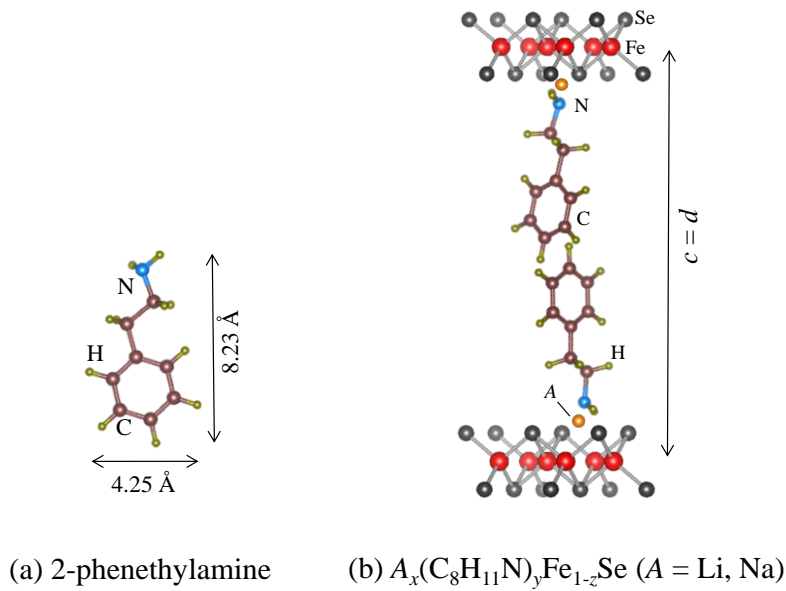

(a) 2-phenethylamine    (b) $A_x(C_8H_{11}N)_yFe_{1-z}Se$ ($A$ = Li, Na)

Fig. 2. Schematic views of (a) 2-phenethylamine (2-PEA) and (b) the possible crystal structure of $A_x(C_8H_{11}N)_yFe_{1-z}Se$ ($A$ = Li, Na).

In this crystal structure, two tendencies characteristic of the FeSe-based intercalation compounds are taken into account: one is that alkali metal ions are located near Se atoms [38] and the other is that lone-pair electrons of N atoms are attracted by alkali metal ions.[30,39] In fact, a very similar structure of intercalants has been observed in phenethylammonium-intercalated $CuCl_4$.[40,41]

Figure 3 shows the temperature dependences of $\chi$ in a magnetic field of 10 Oe on zero-field cooling (ZFC) and on field cooling (FC) for intercalated samples consisting of $A_x(C_8H_{11}N)_yFe_{1-z}Se$ ($A$ = Li, Na) and FeSe. The first and second superconducting transitions are observed at 39 – 44 and 9 K, respectively. By taking into account the powder x-ray diffraction result, it is concluded that the first and second transitions are due to the bulk superconductivity of $A_x(C_8H_{11}N)_yFe_{1-z}Se$ ($A$ = Li, Na) and the non-intercalated region of FeSe, respectively. It is noted that the hysteresis of $\chi$ above $T_c$ is due to magnetic impurities taken into the samples, as in the case of $A_x(C_2H_8N_2)_yFe_{2-z}Se_2$ ($A$ = Li, Na), $Li_x(C_6H_{16}N_2)_yFe_{2-z}Se_2$, and $Li_xFe(Se,Te)$.[4,13,16,18,30]

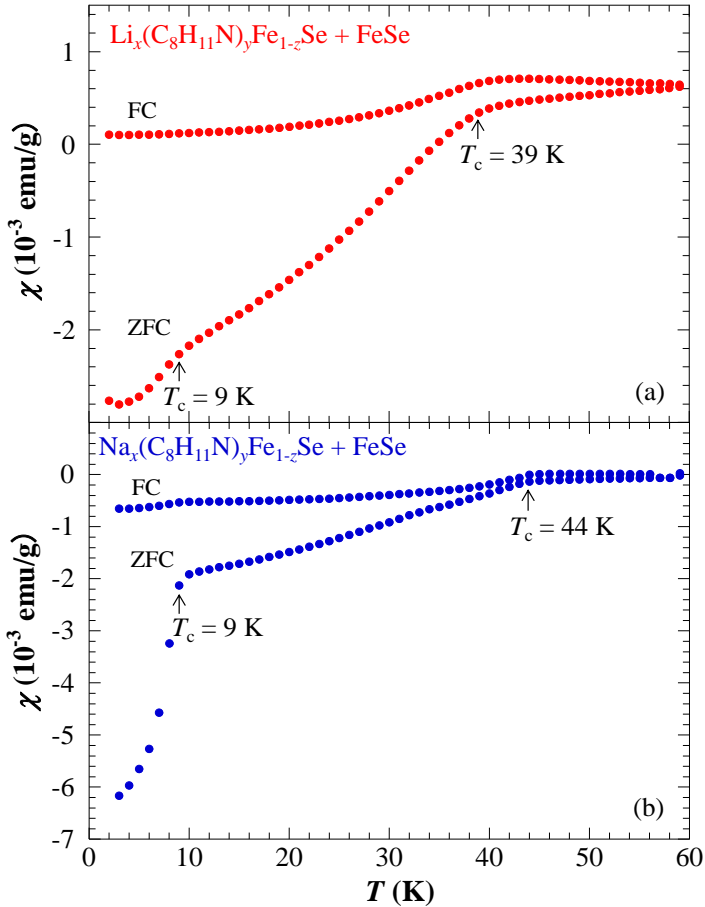

Fig. 3. Temperature dependences of the magnetic susceptibility $\chi$ in a magnetic field of 10 Oe on zero-field cooling (ZFC) and field cooling (FC) for intercalated samples consisting of $A_x(C_8H_{11}N)_yFe_{1-z}Se$ [(a) $A$ = Li, (b) $A$ = Na] and FeSe.

Figure 4 shows the TG curve on heating up to 900 ℃ at a rate of 1 ℃/min for the intercalated sample consisting of $Li_x(C_8H_{11}N)_yFe_{1-z}Se$ and FeSe. Roughly, two steps of mass loss are observed: a 27% loss below 300 ℃ and a large loss above 700 ℃. The first step below ~ 300 ℃ is inferred to be due to the deintercalation or desorption of 2-PEA, because very similar TG curves have been observed for other FeSe-based intercalation compounds such as $Li_x(C_2H_8N_2)_yFe_{2-z}Se_2$ [18] and $Li_x(C_6H_{16}N_2)_yFe_{2-z}Se_2$,[30,39] where the first and second steps are reasonably understood to be due to the deintercalation of organic molecules and the decomposition of FeSe, respectively. Assuming that the composition of the intercalation compound is $x = y = 1$ in $Li_x(C_8H_{11}N)_yFe_{1-z}Se$ based on the nominal composition, the molar ratio of the intercalation compound $Li(C_8H_{11}N)FeSe$ to the host compound FeSe in the intercalated sample is estimated from the mass loss below 300 ℃ to be about 42%. On the other hand, assuming $x = y = 0.5$ due to the less dense packing of the intercalants, the molar ratio of the intercalation compound $Li_{0.5}(C_8H_{11}N)_{0.5}FeSe$ to the host compound FeSe in the intercalated sample is estimated to be about 85%. By taking into account both the powder x-ray diffraction result shown in Fig. 1, that is, peaks due to FeSe are weaker than those due to the Li- and 2-PEA-intercalated compound, and the result for $\chi$ shown in Fig. 3, that is, the change in $\chi$ caused by the superconducting transition due to FeSe is smaller than that due to the Li- and 2-PEA-intercalated compound, the latter seems more plausible than the former, although a detailed analysis will be necessary.

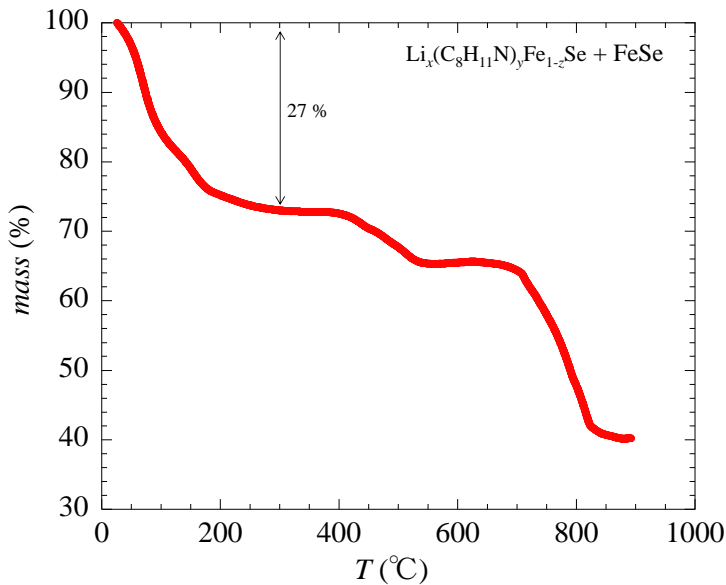

Fig. 4. Thermogravimetric (TG) curve on heating at a rate of 1 ℃/min for the intercalated sample consisting of $Li_x(C_8H_{11}N)_yFe_{1-z}Se$ and FeSe.

Figure 5 shows the relationship between $T_c$ and $d$ for the present samples of $A_x(C_8H_{11}N)_yFe_{1-z}Se$ ($A$ = Li, Na) and a variety of other FeSe-based intercalation superconductors.[1-12,15-18,20,21,30] It is found that $T_c$ vs $d$ is not domic but is saturated at ~ 45 K for $d \geq 9$ Å. This relationship between $T_c$ and $d$ is well understood in terms of the electronic structure calculated by Guterding $et\ al.$[42] with density functional theory. That is, the electronic structure becomes more two-dimensional with increasing $d$, leading to the increase in $T_c$, and it is perfectly two-dimensional at $d = 8 - 10$ Å, leading to the saturation of $T_c$ at $d \geq 8$ Å. According to the random-phase-approximation spin-fluctuation approach by Guterding $et\ al.$,[42] dispersive values of $T_c$ at $d \geq 8$ Å are also understood as being due to the difference in the number of carriers doped into the two-dimensional Fermi surface. They may also be related to the local lattice distortion such as the change in the Se height from the Fe layer.[15,43]

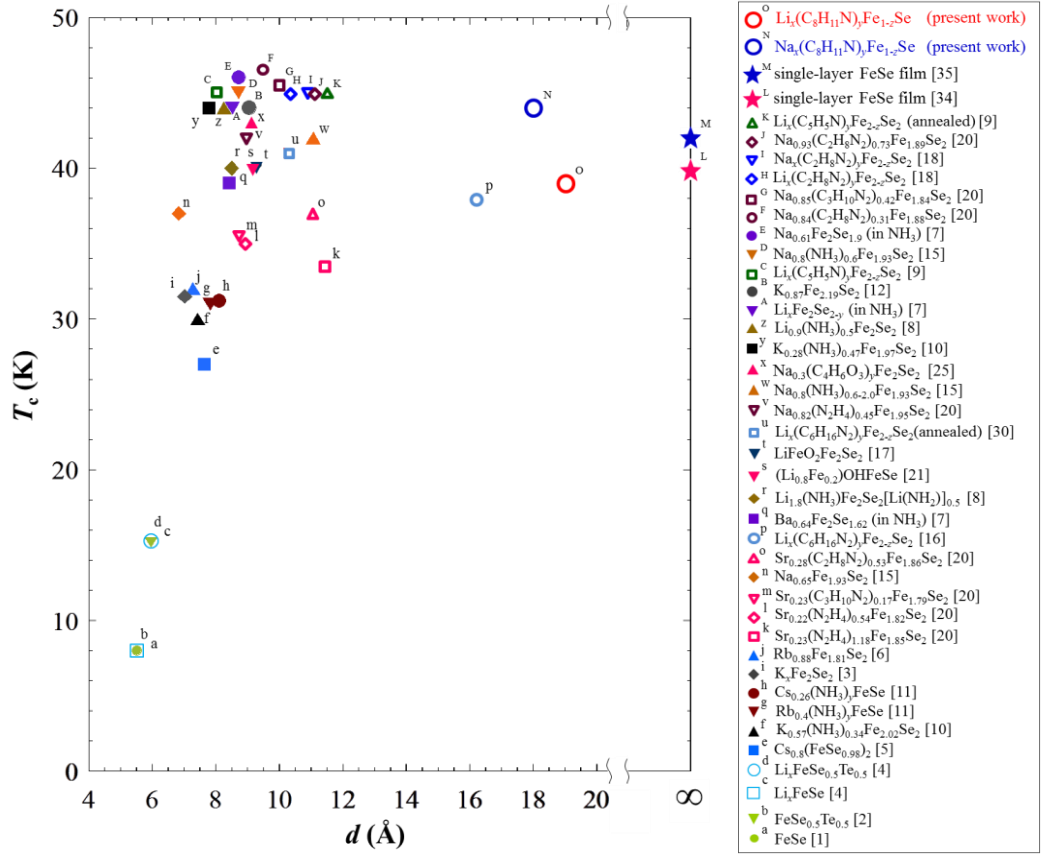

Fig. 5. Relationship between $T_c$ and the interlayer spacing between neighboring Fe layers, $d$, in the FeSe-based intercalation superconductors.[1-12,15-18,20,21,30] Assuming that the $d$ values of single-layer FeSe films are infinite, their $T_c$ values estimated from the resistive measurements are also plotted for reference.[34,35]

Finally, we comment on the implication of the present result in relation to the single-layer FeSe films mentioned above. Considering that the electronic structure of the FeSe-based intercalation superconductors is perfectly two-dimensional at $d \geq 8$ Å, as pointed out by Guterding *et al.*,[42] and that $T_c$ is actually saturated at ~ 45 K for large $d$ values, the electronic structure of the single-layer FeSe films is inferred to be similar to that of the FeSe-based intercalation superconductors with large $d$ values. Therefore, it is reasonable that a resistive superconducting transition has been observed at ~ 40 K in single-layer films.[31,34,35] The opening of a superconducting gap at a temperature as high as 42 - 65 K observed in the STM/STS [31] and ARPES [32,33] measurements is inferred to be due to the superconducting fluctuation, because $T_c^{onset}$, defined as the temperature where the electrical resistivity starts to decrease with decreasing temperature due to the superconducting transition, is above 50 K in the single-layer FeSe films.[31] As a matter of fact, $T_c^{onset}$ due to the superconducting fluctuation of FeSe-based intercalation superconductors such as $A_x(C_2H_8N_2)_yFe_{2-z}Se_2$ ($A$ = Li, Na) is as high as 55 – 57 K.[44] Accordingly, it is concluded that the electronic structure of the single-layer FeSe films is very similar to that of the FeSe-based intercalation superconductors with large $d$ values and that, unfortunately, no further increase in $T_c$ seems to be expected in any FeSe-based superconductors with $d > 19$ Å.

To summarize, we have succeeded in synthesizing new FeSe-based intercalation compounds, $A_x(C_8H_{11}N)_yFe_{1-z}Se$ ($A$ = Li, Na), via the co-intercalation of alkali metals and 2-PEA into FeSe. The $d$ values of $A_x(C_8H_{11}N)_yFe_{1-z}Se$ ($A$ = Li, Na) are 19.04(6) and 18.0(1) Å, respectively, and are the largest among those of the FeSe-based intercalation compounds and are understood to due to the intercalation of two molecules of 2-PEA in series perpendicular to the FeSe layers. It seems that the composition of the Li- and 2-PEA-intercalated compound in the intercalated sample is roughly $Li_{0.5}(C_8H_{11}N)_{0.5}FeSe$. Bulk superconductivity of $A_x(C_8H_{11}N)_yFe_{1-z}Se$ ($A$ = Li, Na) has been observed below 39 and 44 K in the $\chi$ measurements, respectively. It has been found that the relationship between $T_c$ and $d$ in the FeSe-based intercalation superconductors is not domic but $T_c$ is saturated at ~ 45 K for $d \geq 9$ Å. $T_c$ ~ 45 K is comparable to not $T_c^{onset}$ values but mean $T_c$ values of single-layer FeSe films obtained from resistive measurements. Accordingly, it is concluded that the electronic structure of the single-layer FeSe films is very similar to that of the FeSe-based intercalation superconductors with large $d$ values. Unfortunately, no further increase in $T_c$ seems to be expected in any FeSe-based superconductors with $d > 19$ Å.

**Acknowledgments**

This work was supported by JSPS KAKENHI (Grant Numbers 15K13512 and 16K05429). One of the authors (T. H.) was also supported by a Grant-in-Aid for JSPS Fellows (Grant Number 15J00940).

**References**

1) F.-C. Hsu, J.-Y. Luo, K.-W. Yeh, T.-K. Chen, T.-W. Huang, P. M. Wu, Y.-C. Lee, Y.-L. Huang, Y.-Y. Chu, D.-C Yan, and M.-K. Wu, Proc. Natl. Acad. Sci. U.S.A. **105**, 14262 (2008).

2) K.-W. Yeh, T.-W. Huang, Y.-L. Huang, T.-K. Chen, F.-C. Hsu, P. M. Wu, Y.-C. Lee, Y.-Y. Chu, C.-L. Chen, J.-Y. Luo, D.-C. Yan, and M.-K. Wu, Europhys. Lett. **84**, 37002 (2008).

3) J. Guo, S. Jin, G. Wang, S. Wang, K. Zhu, T. Zhou, M. He, and X. Chen, Phys. Rev. B **82**, 180520(R) (2010).

4) H. Abe, T. Noji, M. Kato, and Y. Koike, Physica C **470**, S487 (2010).

5) A. Krzton-Maziopa, Z. Shermadini, E. Pomjakushina, V. Pomjakushin, M. Bendele, A. Amato, R. Khasanov, H. Luetkens, and K. Conder, J. Phys.: Condens. Matter **23**, 052203 (2011).

6) A. F. Wang, J. J. Ying, Y. J. Yan, R. H. Liu, X. G. Luo, Z. Y. Li, X. F. Wang, M. Zhang, G. J. Ye, P. Cheng, Z. J. Xiang, and X. H. Chen, Phys. Rev. B **83** 060512(R) (2011).

7) T. P. Ying, X. L. Chen, G. Wang, S. F. Jin, T. T. Zhou, X. F. Lai, H. Zhang, and W. Y. Wang, Sci. Rep. **2**, 426 (2012).

8) E.-W. Scheidt, V. R. Hathwar, D. Schmitz, A. Dunbar, W. Scherer, F. Mayr, V. Tsurkan, J. Deisenhofer, and A. Loidl, Eur. Phys. J. B **85**, 279 (2012).

9) A. Krzton-Maziopa, E. V. Pomjakushina, V. Y. Pomjakushin, F. von Rohr, A. Schilling, and K. Conder, J. Phys.: Condens. Matter **24**, 382202 (2012).

10) T. Ying, X. Chen, G. Wang, S. Jin, X. Lai, T. Zhou, H. Zhang, S. Shen, and W. Wang, J. Am. Chem. Soc. **135**, 2951 (2013).

11) L. Zheng, M. Izumi, Y. Sakai, R. Egushi, H. Goto, Y. Takabayashi, T. Kambe, T. Onji, S. Araki, T. C. Kobayashi, J. Kim, A. Fujiwara, and Y. Kubozono, Phys. Rev. B **88**, 094521 (2013).

12) A.-m. Zhang, T.-l. Xia, K. Liu, W. Tong, Z.-r. Yang, and Q.-m. Zhang, Sci. Rep. **3**, 1216 (2013).

13) T. Hatakeda, T. Noji, T. Kawamata, M. Kato, and Y. Koike, J. Phys. Soc. Jpn. **82**, 123705 (2013).
14) M. Burrard-Lucas, D. G. Free, S. J. Sedlmaier, J. D. Wright, S. J. Cassidy, Y. Hara, A. J. Corkett, T. Lancaster, P. J. Baker, S. J. Blundell, and S. J. Clarke, Nat. Mater. **12**, 15 (2013).
15) J. Guo, H. Lei, F. Hayashi, and H. Hosono, Nat. Commun. **5**, 4756 (2014).
16) S. Hosono, T. Noji, T. Hatakeda, T. Kawamata, M. Kato, and Y. Koike, J. Phys. Soc. Jpn. **83**, 113704 (2014).
17) X. F. Lu, N. Z. Wang, G. H. Zhang, X. G. Luo, Z. M. Ma, B. Lei, F. Q. Huang, and X. H. Chen, Phys. Phys. Rev. B **89**, 020507(R) (2014).
18) T. Noji, T. Hatakeda, S. Hosono, T. Kawamata, M. Kato, and Y. Koike, Physica C **504**, 8 (2014).
19) S. J. Sedlmaier, S. J. Cassidy, R. G. Morris, M. Drakopoulos, C. Reinhard, S, J. Moorhouse, D. O'Hare, P. Manuel, D. Khalyavin, and S. J. Clarke, J. Am. Chem. Soc. **136**, 630 (2014).
20) F. Hayashi, H. Lei, J. Guo, and H. Hosono, Inorg. Chem. **54**, 3346 (2015).
21) X. F. Lu, N. Z. Wang, H. Wu, Y. P. Wu, D. Zhao, X. Z. Zeng, X. G. Luo, T. Wu, W. Bao, G. H. Zhang, F. Q. Huang, Q. Z. Huang, and X. H. Chen, Nat. Mater. **14**, 325 (2015).
22) K. V. Yusenko, J. Sottmann, H. Emerich, W. A. Crichton, L. Malavasi, and S. Margadonna, Chem. Commun. **51**, 7112 (2015).
23) L. Zheng, X. Miao, Y. Sakai, M. Izumi, H. Goto, S. Nishiyama, E. Uesugi, Y. Kasahara, Y. Iwasa, and Y. Kubozono, Sci. Rep. **5**, 12774 (2015).
24) F. R. Foronda, S. Ghannadzadeh, S. J. Sedlmaier, J. D. Wright, K. Burns, S. J. Cassidy, P. A. Goddard, T. Lancaster, S. J. Clarke, and S. J. Blundell, Phys. Rev. B **92**, 134517 (2015).
25) T. Kajita, T. Kawamata, T. Noji, T. Hatakeda, M. Kato, Y. Koike, and T. Itoh, Physica C **519**, 104 (2015).
26) S.-J. Shen, T.-P. Ying, G. Wang, S.-F. Jin, Z. Han, Z.-P. Lin, and X.-L. Chen, Chin. Phys. B **24**, 117406 (2015).
27) U. Pachmayr, F. Nitsche, H. Luetkens, S. Kamusella, F. Brückner, R. Sarkar, H.-H. Klauss, and D. Johrendt, Angew. Chem. Int. Ed. **54**, 293 (2015).
28) X. Dong, H. Zhou, H. Yang, J. Yuan, K. Jin, F. Zhou, D. Yuan, L. Wei, J. Li, X. Wang, G. Zhang, and Z. Zhao, J. Am. Chem. Soc. **137**, 66 (2015).
29) H. Sun, D. N. Woodruff, S. J. Cassidy, G. M. Allcroft, S. J. Sedlmaier, A. L. Thompson, P. A. Bingham, S. D. Forder, S. Cartenet, N. Mary, S. Ramos, F. R.

Foronda, B. H. Williams, X. Li, S. J. Blundell, and S. J. Clarke, Inorg. Chem. **54**, 1958 (2015).

30) S. Hosono, T. Noji, T. Hatakeda, T. Kawamata, M. Kato, and Y. Koike, J. Phys. Soc. Jpn. **85**, 013702 (2016).

31) Q.-Y. Wang, Z. Li, W.-H. Zhang, Z.-C. Zhang, J.-S. Zhang, W. Li, H. Ding, Y.-B. Ou, P. Deng, K. Chang, J. Wen, C.-L. Song, K. He, J.-F. Jia, S.-H. Ji, Y.-Y. Wang, L.-L. Wang, X. Chen, X.-C. Ma, and Q.-K. Xue, Chin. Phys. Lett. **29**, 037402 (2012).

32) S. Tan, Y. Zhang, M. Xia, Z. Ye, F. Chen, X. Xie, R. Peng, D. Xu, Q. Fan, H. Xu, J. Jiang, T. Zhang, X. Lai, T. Xiang, J. Hu, B. Xie, and D. Feng, Nat. Mater. **12**, 634 (2013).

33) Y. Miyata, K. Nakayama, K. Sugawara, T. Sato, and T. Takahashi, Nat. Mater. **14**, 775 (2015).

34) Y. Sun, W. Zhang, Y. Xing, F. Li, Y. Zhao, Z. Xia, L. Wang, X. Ma, Q.-K. Xue, and J. Wang, Sci. Rep. **4**, 6040 (2014).

35) J. Shiogai, Y. Ito, T. Mitsuhashi, T. Nojima, and A. Tsukazaki, Nat. Phys. **12**, 42 (2016).

36) F. Izumi and K. Momma, Solid State Phenom. **130**, 15 (2007).

37) It is possible for Bragg peaks due to the intercalation compound to be indexed on the basis of the ThCr$_2$Si$_2$-type (*I*4/*mmm*) structure as in the case of $A_x$(C$_2$H$_8$N$_2$)$_y$Fe$_{2-z}$Se$_2$ (*A* = Li, Na) [13,18] and Li$_x$(C$_6$H$_{16}$N$_2$)$_y$Fe$_{2-z}$Se$_2$.[16,30] Here, the PbO-type (*P*4/*nmm*) structure is adopted as in the case of NH$_3$-rich Li$_x$(NH$_3$)$_y$Fe$_{2-z}$Se$_2$ [19] and (Li$_{0.8}$Fe$_{0.2}$)OHFeSe,[21] because the PbO-type structure tends to be plausible in the case that intercalants have no mirror reflection plane parallel to the FeSe layers. In future, a detailed structure analysis of the present intercalation compound is necessary, although the *d* value estimated is independent of the adopted structure.

38) A. M. Alekseeva, O. A. Drozhzhin, K. A. Dosaev, E. V. Antipov, K. V. Zakharov, O. S. Volkova, D. A. Chareev, A. N. Vasiliev, C. Koz, U. Schwarz, H. Rosner, and Y. Grin, Sci. Rep. **6**, 25624 (2016).

39) S. Hosono, T. Noji, T. Hatakeda, T. Kawamata, M. Kato, and Y. Koike, J. Phys. Soc. Jpn. **85,** 104701 (2016).

40) R. D. Willett, Acta Cryst. C**46**, 565 (1990).

41) A. O. Polyakov, A. H. Arkenbout, J. Baas, G. R. Blake, A. Meetsma, A. Caretta, P. H. M. van Loosdrecht, and T. T. M. Palstra, Chem. Mater. **24**, 133 (2012).

42) D. Guterding, H. O. Jeschke, P. J. Hirschfeld, and R. Valenti, Phys. Rev. B **91**,

041112(R) (2015).

43) E. Paris, L. Simonelli, T. Wakita, C. Marini, J.-H. Lee, W. Olszewski, K. Terashima, T. Kakuto, N. Nishimoto, T. Kimura, K. Kudo, T. Kambe, M. Nohara, T. Yokoya, and N. L. Saini, Sci. Rep. **6**, 27646 (2016).

44) T. Hatakeda, T. Noji, S. Hosono, T. Kawamata, M. Kato, and Y. Koike, J. Phys.: Conf. Ser. **568**, 022032 (2014).